\documentclass[pra, showpacs, floatfix, preprint, endfloats*]{revtex4}
\usepackage{mathrsfs}
\usepackage{amsmath}
\usepackage{amssymb}
\usepackage{revsymb}
\usepackage{graphicx}
\usepackage{mathrsfs}

\begin{document}
\title{Enhancement of vacuum polarization effects in a plasma: correct after online version appears}
\author{A. \surname{Di Piazza}}
  \email{dipiazza@mpi-hd.mpg.de}
  \author{K. Z. \surname{Hatsagortsyan}}
  \email{k.hatsagortsyan@mpi-hd.mpg.de}
  \author{C. H. \surname{Keitel}}
  \email{keitel@mpi-hd.mpg.de}
  \affiliation{Max-Planck-Institut f\"ur Kernphysik,
    Saupfercheckweg 1, D-69117 Heidelberg, Germany}

\date{\today}

\begin{abstract}
The dispersive effects of vacuum polarization on the propagation of a strong circularly polarized electromagnetic wave through a cold collisional plasma are studied analytically. It is found that, due to the singular dielectric features of the plasma, the vacuum effects on the wave propagation in a plasma are qualitatively different and much larger than those in pure vacuum in the regime when the frequency of the propagating wave approaches the plasma frequency. A possible experimental setup to detect these effects in plasma is described.

 
\end{abstract}
 
\maketitle

\section{Introduction}
The fast development of laser technology in the last years has enabled novel applications of lasers in different areas \cite{Salamin_2006,Mourou_2006}. Terawatt and petawatt table-top lasers are already employed, for example, to create new short-wavelength sources up to x-rays and $\gamma$-rays \cite{x-rays}, to accelerate particles with enormous acceleration rates (even of the order of $\text{GeV/cm}$ in the case of electrons) \cite{Laccel} and as drivers for nuclear fusion \cite{Fusion}. Strong optical laser pulses with intensities of order of $10^{22}\;\text{W/cm$^2$}$ have been already produced in laboratories \cite{Bahk_2004} and intensities of order of $10^{24}\text{-}10^{26}\;\text{W/cm$^2$}$ are envisaged \cite{Tajima_2002}. In the presence of such strong electromagnetic fields it is becoming feasible to probe the nonlinear properties of vacuum as predicted by quantum electrodynamics (see the book in Ref. \cite{Dittrich_b_2000} and the references therein). In fact, starting from the early works by Delbr\"uck \cite{Delbrueck}, Heisenberg and Euler \cite{Heisenberg_1936} and Weisskopf \cite{Weisskopf_1936}, numerous theoretical papers have been devoted to investigate how the presence of strong electromagnetic fields modifies the dielectric properties of the vacuum \cite{Schwinger_1951,Vac_Pol_Static,Becker_1975,Baier_1975,Aleksandrov_1986}. The typical electric (magnetic) field strength at which these effects are predicted to become apparent is given by $E_{cr}=m_e^2c^3/\hbar e=1.3\times 10^{16}\;\text{V/cm}$ ($B_{cr}=m_e^2c^3/\hbar e=4.4\times 10^{13}\;\text{G}$) corresponding to the huge intensity: $I_{cr}=cE^2_{cr}/4\pi=4.6\times 10^{29}\;\text{W/cm$^2$}$. In the above expressions $-e$ and $m_e$ are the negative charge and the mass of the electron, respectively. From a theoretical point of view, three different classes of electromagnetic fields have been considered as candidates able to ``polarize'' the vacuum: Coulomb fields of highly charged nuclei \cite{Delbrueck}, static magnetic fields \cite{Vac_Pol_Static} and laser fields \cite{Becker_1975,Baier_1975,Aleksandrov_1986,Di_Piazza_2005,Heinzl_2006,Lundstroem_2006,Di_Piazza_2006}. So far, an experimental confirmation of the theoretical predictions has been achieved only for the first class with the observation of Delbr\"uck scattering \cite{Experiment_Delb} and the related process of $\gamma$-photon splitting in the field of a heavy nucleus \cite{Experiment_PS}. On the other hand, the PVLAS (Polarizzazione del Vuoto con Laser) experiment is devoted to measure the extremely small ellipticity acquired by a linearly polarized probe laser after passing repeatedly through a vacuum region with an applied static uniform magnetic field of strength $5.5\times 10^4\;\text{G}$ \cite{PVLAS}. Experimental results have already been reported in Ref. \cite{Zavattini_2006} but they cannot be explained as a nonlinear quantum electrodynamics effect but as an effect due to the possible conversion of a photon into a pseudoscalar particle, called axion. Finally, in the experiments so far performed no interaction has been detected between strong laser beams in vacuum \cite{Moulin}. Nevertheless, recent estimations of vacuum polarization effects (VPEs) in laser-laser collisions, as in Ref. \cite{Heinzl_2006} in the geometrical optics limit and in Ref. \cite{Di_Piazza_2006} with the inclusion of diffractive effects, have shown a feasibility to observe VPEs in the near future. 

When a strong laser wave, able to polarize the vacuum, encounters matter, a plasma is immediately created. Already at intensities above $10^{18}\;\text{W/cm$^2$}$ for optical frequencies this plasma is relativistic and drastically modifies the laser wave evolution. Laser self-focusing, self-channeling and plasma self-induced transparency are only a few of the numerous effects arising during the propagation of a laser field through a relativistic plasma (see the book in Ref. \cite{Borovski_b_2003}, the recent reviews in Refs. \cite{Umstadter_2003,Marklund_2006} and the references therein). From another side, super-strong short-wavelength electromagnetic radiation fields approaching the value of the critical field have been predicted via the interaction of a strong laser pulse with an inhomogeneous plasma \cite{Bulanov}. In this situation, VPEs could significantly modify the dynamics both of the plasma and of the laser field. The influence of the absorptive features of the vacuum polarization, namely the possibility of pair creation, on the propagation of a strong electromagnetic wave through an electron-positron plasma has been considered in Ref. \cite{Bulanov_2005} where the dispersive effects of vacuum polarization have been neglected. The related back-reaction effects in the pair production process in crossed laser beams have been considered in Ref. \cite{non-Markovian} and non-Markovian effects in the pair production kinetics have been found.
The dispersive VPEs can significantly modify the radiation propagation in a plasma as it has been shown in Refs. \cite{Gnedin_1978,Meszaros_1978} and more recently in Refs. \cite{Stenflo_2005,Brodin_2006,Marklund_Nonlin,Lundin_2006} (see also the review in Ref. \cite{Marklund_2006}). The authors refer to astrophysical scenarios where VPEs are due to a strong static magnetic field and transform the dispersion equation of a probe wave in such a way to allow the propagation of a new low-frequency mode in plasma. It has been shown also that in nonlinear regime a shock wave can be formed in magnetized plasmas due to the VPEs which could be important for the evolution of supernova remnants \cite{Marklund_Nonlin}. Also, vacuum polarization corrections to wave propagation in unmagnetized as well as in magnetized plasma have been investigated in Ref. \cite{Lundin_2006}. Another manifestation of the VPEs in plasma, namely, the photon-photon scattering in a laser-produced channel in overdense plasma has been considered in Ref. \cite{Shen_2003}. The plasma channel traps the laser field providing high radiation field intensity on relatively long distances. 

In this paper we investigate the role of dispersive VPEs in the strong laser wave propagation in plasma in a new regime of physical parameters.
In fact, the problem of the propagation of a strong laser wave in a relativistic plasma has been studied for the first time in the seminal work in Ref.  \cite{Akhiezer_1956} by Akhiezer and Polovin. In the paper in Ref. \cite{Akhiezer_1956} an exact analytical solution of this problem was found in the case of the propagation of a plane wave with circular polarization. The solution obtained in Ref. \cite{Akhiezer_1956} holds under ideal assumptions like the plane-wave one but it has provided a qualitative understanding of what happens in a relativistic plasma in more complicated situations. Up to now the Akhiezer-Polovin solution remains the only exact solvable model for the strong laser wave propagation in plasma, while nowadays, the quantitative theoretical study of relativistic plasma dynamics in realistic conditions is mostly performed via numerical solutions.  In view of the ever increasing available laser intensities, we reconsider in the present paper the Akhiezer-Polovin problem by taking into account the VPEs. Since the critical intensity $I_{cr}$ remains many orders of magnitude larger than the available ones, a first-order perturbative solution of the problem in VPEs will be obtained. The solution shows a large enhancement of the VPEs in plasma with respect to those predicted in pure vacuum when the frequency of the probe field (that in our case is also the field that polarizes the vacuum) approaches the plasma frequency. A decisive role in this sense is played by the singular dielectric behaviour of the plasma near the plasma frequency. In contrast to this, the plasma was assumed in Refs. \cite{Gnedin_1978,Meszaros_1978,Stenflo_2005,Brodin_2006} to be tenuous that is, the frequency of the radiation propagating in the plasma is much larger than the plasma frequency. In that regime the enhancement of the VPEs we consider here, is completely suppressed. It must also be stressed that, despite quantum electrodynamics is a well established theory from an experimental point of view, dispersive VPEs induced by laser fields have not yet been observed. The reason is essentially that the observation of these effects requires very intense lasers beams. In the following we propose a possible experimental implementation of the ideas previously mentioned in order to exploit the plasma enhancement of VPEs to measure for the first time VPEs themselves.

The paper is organized as follows. In the next Section the theoretical model is described. In Sec. \ref{Results} exact analytical results are obtained in the case of a collisionless plasma and discussed with particular emphasis on the corrections introduced by the VPEs on the strong wave propagation. In Sec. \ref{Setup} a possible experimental setup to measure the VPEs in plasma is described. Finally, in Sec. \ref{Collisions} the effects of electron-ion collisions are taken into account analytically by a perturbative approach and the corresponding modifications required in the proposed setup are discussed. The main conclusions are summarized in Sec. \ref{Conclusions}. In the following, natural units with $\hbar=c=1$ will be used throughout.
%
%
\section{Theoretical model}
In Ref. \cite{Akhiezer_1956} Akhiezer and Polovin solved exactly the problem of the propagation of a circularly polarized plane wave through a cold relativistic collisionless plasma. In their calculations the plasma was described as a one-component relativistic electron fluid moving in the presence of a fixed ion background. In the following we consider the analogous problem of the propagation of a strong electromagnetic field with circular polarization through a cold plasma but including the VPEs induced by the strong field itself. In contrast to Ref. \cite{Akhiezer_1956}, we will take into account the motion of the ions because, for example, a proton moves relativistically in the presence of an optical laser field with intensity we are interested in of order of $10^{23}\text{-}10^{24}\;\text{W/cm$^2$}$. The approximation of cold plasma is well justified because the thermal motion can be neglected with respect to the highly relativistic motion driven by the external strong field. Finally, we also include the effects of the electron-ion collisions because they can play a role in the frequency region we are interested in, that is close to the plasma frequency. The equations that describe the previous physical scenario are essentially Maxwell equations in the presence of two real charged fluids with their fluids equations of motion, and of an additional ``vacuum fluid'' represented by the four-current $(\rho_{\text{vac}}(\mathbf{r},t),\mathbf{J}_{\text{vac}}(\mathbf{r},t))$ analyzed below:
\begin{align}
\label{Div_D}
\boldsymbol{\partial}\cdot \mathbf{E}&=-e(N_e-ZN_i)+\rho_{\text{vac}},\\
\boldsymbol{\partial}\cdot \mathbf{B}&=0,\\
\boldsymbol{\partial}\times\mathbf{E}+\partial_t\mathbf{B}&=\mathbf{0},\\
\label{Rot_H}
\boldsymbol{\partial}\times\mathbf{B}-\partial_t\mathbf{E}&=-e(N_e\mathbf{v}_e-ZN_i\mathbf{v}_i)+\mathbf{J}_{\text{vac}},\\
\label{Cont_e}
\partial_t N_e+\boldsymbol{\partial}\cdot (N_e\mathbf{v}_e)&=0,\\
\label{Cont_p}
\partial_t N_i+\boldsymbol{\partial}\cdot (N_i\mathbf{v}_i)&=0,\\
\label{Eq_mot_e}
\partial_t \mathbf{p}_e+(\mathbf{v}_e\cdot\boldsymbol{\partial})\mathbf{p}_e&=-e(\mathbf{E}+\mathbf{v}_e\times\mathbf{B})-\nu_{ei} m_r(\gamma_e\mathbf{v}_e-\gamma_i\mathbf{v}_i),\\
\label{Eq_mot_p}
\partial_t \mathbf{p}_i+(\mathbf{v}_i\cdot\boldsymbol{\partial})\mathbf{p}_i&=Ze(\mathbf{E}+\mathbf{v}_i\times\mathbf{B})-\nu_{ie} m_r(\gamma_i\mathbf{v}_i-\gamma_e\mathbf{v}_e).
\end{align}
It is more convenient to start discussing Eqs. (\ref{Cont_e}) and (\ref{Cont_p}). They are the continuity equations of the electron and ion fluids having densities $N_{\lambda}(\mathbf{r},t)$ and currents $N_{\lambda}(\mathbf{r},t)\mathbf{v}_{\lambda}(\mathbf{r},t)$, respectively: in these expressions and in the following the index $\lambda\in\{e,i\}$ indicates the two particles species at hand, i. e. electrons and ions. The ion mass and charge are $m_i$ and $Ze$, respectively. Also, Eqs. (\ref{Eq_mot_e}) and (\ref{Eq_mot_p}) are the equations of motions of the two fluids in the presence of the electromagnetic field $\mathbf{E}(\mathbf{r},t)$ and $\mathbf{B}(\mathbf{r},t)$ and by accounting phenomenologically for the collisions between the two fluids. The momenta $\mathbf{p}_{\lambda}(\mathbf{r},t)$ are connected with the velocities $\mathbf{v}_{\lambda}(\mathbf{r},t)$ through the usual relativistic relation $\mathbf{p}_{\lambda}(\mathbf{r},t)=m_{\lambda}\gamma_{\lambda}(\mathbf{r},t)\mathbf{v}_{\lambda}(\mathbf{r},t)$ with $\gamma_{\lambda}(\mathbf{r},t)=(1-v^2_{\lambda}(\mathbf{r},t))^{-1/2}$ being the relativistic Lorentz factors. The expressions of the relativistic dissipative terms in Eqs. (\ref{Eq_mot_e}) and (\ref{Eq_mot_p}) can be found in Ref. \cite{Gedalin_1996}. In these terms $m_r=m_e m_i/(m_e+m_i)$ is the reduced mass and $\nu_{ei}$ and $\nu_{ie}$ are the two effective electron-ion collision frequencies that will be defined and discussed in Sec. \ref{Collisions}. In the cold-plasma limit we are working in, the equations of the momenta $\mathbf{p}_{\lambda}(\mathbf{r},t)$ describe completely the motion of the two fluids.

The electromagnetic field evolution in the presence of the electron and ion fluids is described by Maxwell equations (\ref{Div_D})-(\ref{Rot_H}). In these equations the VPEs are included in the auxiliary vacuum four-current ($\rho_{\text{vac}}(\mathbf{r},t),\mathbf{J}_{\text{vac}}(\mathbf{r},t))$ appearing in Eqs. (\ref{Div_D}) and (\ref{Rot_H}) that is given by
\begin{align}
\label{rho_v}
\rho_{\text{vac}}&=-\frac{4\alpha^2}{45m_e^4}\boldsymbol{\nabla}\cdot\left[2(E^2-B^2)\mathbf{E}+7(\mathbf{E}\cdot\mathbf{B})\mathbf{B}\right],\\
\label{J_v}
\mathbf{J}_{\text{vac}}&=-\frac{4\alpha^2}{45m_e^4}\left\{\boldsymbol{\nabla}\times\left[2(E^2-B^2)\mathbf{B}-7(\mathbf{E}\cdot\mathbf{B})\mathbf{E}\right]-\partial_t\left[2(E^2-B^2)\mathbf{E}+7(\mathbf{E}\cdot\mathbf{B})\mathbf{B}\right]\right\}
\end{align}
where $\alpha=e^2/4\pi$ is the fine-structure constant. Despite the vacuum four-current is not made of real particles (only the electromagnetic field appears in it) it fulfills automatically the continuity equation. Maxwell equations (\ref{Div_D})-(\ref{Rot_H}) with the vacuum four-current $(\rho_{\text{vac}}(\mathbf{r},t),\mathbf{J}_{\text{vac}}(\mathbf{r},t))$ can be obtained by applying the variational method to the Lagrangian density $\mathscr{L}_{\text{field}}+\mathscr{L}_{\text{int}}$ where the field-matter interaction Lagrangian density $\mathscr{L}_{\text{int}}$ has the usual expression (see, e. g. Ref. \cite{Landau_b_2_1975}). But, the field Lagrangian density $\mathscr{L}_{\text{field}}$ is not simply the Maxwell one $(E^2-B^2)/2$ but it is given by
\begin{equation}
\label{L_eff}
\mathscr{L}_{\text{field}}=\frac{1}{2}(E^2-B^2)+\frac{2\alpha^2}{45m_e^4}\left[(E^2-B^2)^2+7(\mathbf{E}\cdot\mathbf{B})^2\right]
\end{equation}
i. e. by the Euler-Heisenberg effective Lagrangian density that takes into account the VPEs at the lowest order in the limits of small field frequencies $\omega$ with respect to the electron mass and of small field amplitudes with respect to the critical fields $E_{cr}$ and $B_{cr}$ \cite{Heisenberg_1936}. In general, in the presence of a plasma with density $N$ and temperature $T$, the use of the effective Lagrangian density in Eq. (\ref{L_eff}) is allowed if the frequency $\omega$ is also much smaller than $k_BT$ with $k_B$ the Boltzmann constant and $N^{1/3}$ \cite{Elmfors_1995}. Also, in the presence of particle backgrounds the Euler-Heisenberg effective Lagrangian density (\ref{L_eff}) has to be corrected by adding the so-called effective Lagrangian density at finite temperature and chemical potential \cite{Kapusta_b_1989,Cangemi_1996,Dittrich_b_2000}. It turns out that the finite-temperature two-loop contribution to the effective Lagrangian density is dominant with respect to the one-loop contribution at temperatures such that $k_BT\ll m_e$ \cite{Gies_2000}. In fact, while the one-loop corrections are exponentially damped by a factor $\exp(-m_e/k_BT)$, the two-loop ones scale as $(k_BT/m_e)^4$. It can be shown that these terms are negligible with respect to the Euler-Heisenberg effective Lagrangian density if $k_BT\ll \sqrt{eE}, \sqrt{eB}$ \cite{Footnote_1}. In the following, we will assume that all the previous conditions are fulfilled. Before concluding we mention that also the spontaneous electron-positron pair creation from vacuum has been neglected in Eqs. (\ref{Div_D})-(\ref{Eq_mot_p}). This approximation is very well justified at field amplitudes much smaller than the critical field not only at zero-temperature but also at finite temperatures such that $k_BT\ll m_e$. This is because the one-loop thermal corrections to the imaginary part of the effective Lagrangian density vanish and, as it is shown in Ref. \cite{Gies_2000} for a purely electric field, the imaginary part of the two-loop thermal effective Lagrangian density is also suppressed by the exponential factor $\exp(-\pi E_{cr}/E)$ like the imaginary part of the Euler-Heisenberg effective Lagrangian density.
%
%
\section{A simplified analysis of VPEs for the strong wave propagation in plasma: collisionless case}
\label{Results}
In this Section we study how the presence of VPEs influences the propagation of a strong wave through the plasma. For definites, we assume that the wave electric field amplitude and frequency are  $E$ and $\omega$, respectively. In general, two small parameters characterize our problem. One small parameter is connected with the VPEs: $\delta_{\text{vac}} \equiv (\alpha/45\pi) (E/E_{cr})^2$. We remind that the effective Lagrangian density in Eq. (\ref{L_eff}) is correct up to first order in this parameter. The second small parameter, that we indicate as $\delta_{\nu}$, is connected with the electron-ion collisions. It is determined by the ratio between the collision frequencies $\nu_{ei},\nu_{ie}$ and the laser frequency $\omega$ and it is small because the particles motion is driven by a strong laser field. The inclusion of collisional effects makes the analytical solution of Eqs. (\ref{Div_D})-(\ref{Eq_mot_p}) technically more involved. Since the conclusions of the paper will not be changed, we consider first the easier case of a collisionless plasma and we elucidate the physical mechanism that leads to the enhancement of VPEs. Then, we will discuss in detail the effects of collisions among electrons and ions in Sec. \ref{Collisions}.
%
%
\subsection{Exact analytical solution of the plasma and wave equations and discussion of VPEs}

We study here the propagation of a transverse monochromatic electromagnetic wave along the positive $z$ axis with frequency $\omega$ and amplitude $E$ and with circular polarization. If we set the collision frequencies $\nu_{ei}$ and $\nu_{ie}$ equal to zero in Eqs. (\ref{Eq_mot_e}) and (\ref{Eq_mot_p}) the whole system of Eqs. (\ref{Div_D})-(\ref{Eq_mot_p}) can be solved exactly even including VPEs. Actually, we look for a solution of Eqs. (\ref{Div_D})-(\ref{Eq_mot_p}) formally identical to the Akhiezer-Polovin one:
\begin{align}
N_e(\phi)&=ZN_i(\phi)\equiv N_0,\\
\mathbf{p}_e(\phi)&=-\frac{\mathbf{p}_i(\phi)}{Z}=-\frac{eE}{\omega}(\hat{\mathbf{x}}\cos \phi+\hat{\mathbf{y}}\sin \phi),\\
\label{E_fin}
\mathbf{E}(\phi)&=E(\hat{\mathbf{x}}\sin\phi-\hat{\mathbf{y}}\cos\phi),\\
\label{B_fin}
\mathbf{B}(\phi)&=nE(\hat{\mathbf{x}}\cos \phi+\hat{\mathbf{y}}\sin \phi)
\end{align}
where we have introduced the wave phase $\phi=kz-\omega t=\omega(nz-t)$ with $n=k/\omega$ being the wave refractive index. As in Akhiezer-Polovin case, we have assumed the plasma to remain always locally neutral. This approximation is justified for laser pulses circularly polarized [as that considered in Eqs. (\ref{E_fin}) and (\ref{B_fin})] and much longer than the plasma characteristic time $\sim 2\pi/\omega_{pl,0}$ needed to restore quasineutrality [see Eq. (\ref{omega_pl_0}) below for the definition of the \emph{effective} plasma frequency $\omega_{pl,0}$]. In the following we will assume that this is the case. In particular, we will be interested in the regime $\omega\gtrsim\omega_{pl,0}$; then we will implicitly consider laser pulses lasting many laser periods. The above solution has the property that the two electromagnetic invariants that enter the vacuum polarization four-current are $E^2(\phi)-B^2(\phi)=(1-n^2)E^2=\text{const.}$ and $\mathbf{E}(\phi)\cdot\mathbf{B}(\phi)=0$, respectively. In this way, by plugging the above expressions into Eqs. (\ref{Div_D})-(\ref{Eq_mot_p}) one easily obtains that they solve those equations if
\begin{equation}
\label{n_0}
\begin{split}
n&\equiv n_0=\sqrt{1-\frac{1}{\omega^2}\left(\frac{\omega^2_{pl,e}}{\gamma_{0,e}}+\frac{\omega^2_{pl,i}}{\gamma_{0,i}}\right)+\frac{2\alpha}{45\pi}\frac{E^2}{E^2_{cr}}\frac{1}{\omega^4}\left(\frac{\omega^2_{pl,e}}{\gamma_{0,e}}+\frac{\omega^2_{pl,i}}{\gamma_{0,i}}\right)^2}\\
&=\sqrt{n_{pl,0}^2+\frac{2\alpha}{45\pi}\frac{E^2}{E^2_{cr}}(1-n_{pl,0}^2)^2}
\end{split}
\end{equation}
where we have introduced the non-relativistic plasma frequencies $\omega^2_{pl,\lambda}=Z_{\lambda}N_0e^2/m_{\lambda}$ with $Z_e=1$ and $Z_i=Z$, the relativistic Lorentz factors $\gamma_{0,\lambda}=\sqrt{1+Z^2_{\lambda}e^2E^2/\omega^2m^2_{\lambda}}$ and the wave refractive index in the absence of VPEs
\begin{equation}
\label{n_pl}
n_{pl,0}=\sqrt{1-\frac{1}{\omega^2}\left(\frac{\omega^2_{pl,e}}{\gamma_{0,e}}+\frac{\omega^2_{pl,i}}{\gamma_{0,i}}\right)}=\sqrt{1-\frac{\omega^2_{pl,0}}{\omega^2}}
\end{equation}
with the \emph{effective} plasma frequency being
\begin{equation}
\label{omega_pl_0}
\omega_{pl,0}\equiv\sqrt{\frac{\omega^2_{pl,e}}{\gamma_{0,e}}+\frac{\omega^2_{pl,i}}{\gamma_{0,i}}}.
\end{equation}
The above solution describes exactly the wave propagation through the plasma by including the VPEs. As it is clear from Eq. (\ref{n_0}), the VPEs modify the wave refractive index. We observe that since we are interested in wave frequencies close to the effective plasma frequency $\omega_{pl,0}$ we are not allowed for the moment to expand further the square root in Eq. (\ref{n_0}) with respect to VPEs even if we have neglected from the beginning terms proportional to $\delta^2_{\text{vac}}$ with respect to those of order of unity. In this respect, it is interesting to note that, although we have considered the VPEs up to first order in $\delta_{\text{vac}}$ in the Maxwell equations (\ref{Div_D}) and (\ref{Rot_H}), our final result contains vacuum terms to all orders because they are compared with quantities that in principle can be much smaller than unity. Concerning the influence of VPEs on the refractive index $n_0$ some comments are in order. First, we observe that the VPEs do not result in a simple scaling of the plasma frequency. Moreover, as in pure vacuum, the VPEs imply an increase of the wave refractive index and then a decrease of its speed. One can understand intuitively this fact by noting that including the VPEs means to consider the possibility that during its propagation a photon can ``transform'' into a virtual electron-positron pair that then annihilates giving back the initial photon. It is sensible that due to this ``transformation'' into a pair of massive particles the photon is slowed down because the effective speed of the electron-positron pair is less than the speed of light. Finally, we point out that at small plasma densities, i. e. if $n_{pl,0}\to 1$, the VPEs vanish. On one hand, this corresponds to the fact that in vacuum a plane wave does not polarize the vacuum itself \cite{Dittrich_b_2000}. On the other hand, this implies that in the situation considered here, the VPEs on the refractive index cannot be independent of the plasma quantities, in particular of its density. From this point of view we can conclude that the polarization of the vacuum arises here due to the essential modification of the strong wave by the electromagnetic field produced by the electron and ion currents in plasma driven by the strong wave itself. 

Equation (\ref{n_0}) already contains the indication that an enhancement of VPEs compared with those in vacuum can arise when the frequency of the laser field approaches the plasma frequency. Thus, in the limit $n_{pl,0} \rightarrow 0$, the refractive index is mainly determined by the VPEs: $n_0\to \sqrt{2\delta_{\text{vac}}}$ and it scales as the square root of the small parameter of the vacuum correction $\delta_{\text{vac}}$. Meanwhile in vacuum, the vacuum polarization correction to the refractive index is always proportional to $\delta_{\text{vac}}$. Of course, this limiting case has little physical implication because of the strong reflection of the laser wave from plasma and its absorption in plasma. Nevertheless, the mentioned tendency of enhancement of VPEs remains in the regime when the laser frequency is close to the plasma frequency but when the wave propagation in plasma still takes place. This is discussed in the next Section.
%
%
\section{A possible experimental setup to detect the VPEs in plasma}
\label{Setup}
As we have seen in the previous Section, the VPEs affect the refractive index of the strong wave. Then, in order to detect these effects we need to make at least two waves to interfere. Nevertheless, due to the nonlinearity of the plasma equations, the superposition principle does not hold in the plasma. For this reason we consider the experimental setup schematically described in Fig. 1. A laser beam with intensity $I_0$ and frequency $\omega$ is linearly polarized along the $\hat{\mathbf{x}}$ axis and it is split into two waves with intensities $I_{0,1}$ and $I_{0,2}$ such that $I_0=I_{0,1}+I_{0,2}$. Each wave passes through a quarter-wave plate and becomes circularly polarized but in such a way that the two electric fields rotate in opposite directions. Then, they enter two different plasmas with densities $N_{0,1}$ and $N_{0,2}$. In the approximations discussed in the previous Sections we can write the electric fields of the two waves as in Eq. (\ref{E_fin}) by distinguishing them by means of the index $j=1,2$ and by accounting for the fact that they rotate in opposite directions. We point out here that the electric field amplitude $E$ appearing in Eqs. (\ref{E_fin}) and (\ref{n_0}) is the electric field amplitude of the wave \emph{in the plasma}. Experimentally the laser wave is produced in vacuum and then it enters the plasma. In the case of normal incidence the laser electric field amplitude is connected with the corresponding quantity in vacuum $E_0$ by the formula $E=2E_0/(1+n_0)$ \cite{Landau_b_8_1984} and then the two electric field amplitudes $E_j$ in the plasmas are given by
\begin{equation}
\label{T}
E_j=\frac{2}{1+n_{0,j}}\sqrt{4\pi I_{0,j}}
\end{equation}
with $n_{0,j}$ being the refractive indices of the two plasmas. In this respect, since for each $j$ the quantity $n_{0,j}$ also depends on $E_j$, then both $n_{0,j}$ and $E_j$ must be determined consistently from Eqs. (\ref{n_0}) and (\ref{T}) once the intensity in vacuum $I_{0,j}$ and the density $N_{0,j}$ are given. Now, if after a distance $z$ the two waves exit the plasmas with amplitudes $E'_{0,j}$ and are made to interfere, the total electric field $\mathbf{E}'_0(\phi)$ can be written as
\begin{equation}
\label{E}
\mathbf{E}'_0(\phi)=\text{Re}\left\{e^{i\Phi(\phi)}\left[E'_{0,1}(\hat{\mathbf{y}}+i\hat{\mathbf{x}})+E'_{0,2}(\hat{\mathbf{y}}-i\hat{\mathbf{x}})e^{i2\Delta\phi(z)}\right]\right\}
\end{equation}
where $\Phi(\phi)$ is a common phase and $2\Delta\phi(z)$ is the total phase difference between the two waves. The quantity $\Delta\phi(z)$ can be written as
\begin{equation}
\label{Delta_phi}
\Delta\phi(z)=\frac{1}{2}\omega z\frac{n_{pl,0,2}^2-n_{pl,0,1}^2+\frac{2\alpha}{45\pi}\frac{E_2^2}{E^2_{cr}}(1-n^2_{pl,0,2})^2-\frac{2\alpha}{45\pi}\frac{E_1^2}{E^2_{cr}}(1-n^2_{pl,0,1})^2}{n_{0,1}+n_{0,2}}.
\end{equation}
The wave with electric field (\ref{E}) is elliptically polarized but due to the phase difference in the second term the main axes of the ellipse are rotated with respect to the $\hat{\mathbf{x}}$ and $\hat{\mathbf{y}}$ axes by the angle $\Delta \phi(z)$. This rotation angle is the quantity to be measured in order to detect the VPEs. Equation (\ref{Delta_phi}) clearly shows the possibility with respect to pure vacuum of enhancing the VPEs in plasma by exploiting the singular behaviour of the plasma refractive index for laser frequencies close to the effective plasma frequency. In fact, the typical expression of the analogous phase difference in vacuum corresponding to $2\Delta\phi(z)$ is given by $2\Delta\phi_{\text{vac}}(z)\sim(\alpha/45\pi)(E/E_{cr})^2\omega z$. Then, if in Eq. (\ref{Delta_phi}) $n_{0,j}\ll 1$, i. e. if $\omega\gtrsim\omega_{pl,0,j}$, the polarization rotation angle in plasma due to VPEs is much larger than $\Delta\phi_{\text{vac}}(z)$.

Nevertheless, we have to point out that while in pure vacuum the \emph{total} phase difference is already proportional to the VPEs, instead in plasma the phase difference contains a plasma contribution. Then, the problem here is to isolate the VPEs in a measurable way. To achieve this we require the two waves to have the same zero-order refractive index $n_{pl,0}(N_0,I_0)$ (for the sake of clarity, we have explicitly indicated the dependence of the refractive index on the plasma density and on the wave intensity in vacuum). Although, it is desirable that the laser intensities of these two beams are differing as much as possible to have a larger difference in the refractive index due to VPEs [see Eq. (\ref{Delta_phi})]. We can achieve our goal by choosing different densities $N_{0,1}$ and $N_{0,2}$ for the two plasmas. Thus, for the fixed densities $N_{0,1}$ and $N_{0,2}$ as well as the laser frequency $\omega$, the laser intensities $I_{0,1}$ and $I_{0,2}$ have to be adjusted in such a way [see Eq. (\ref{n_pl})] that: $n_{pl,0,1}(N_{0,1},I_{0,1})=n_{pl,0,2}(N_{0,2},I_{0,2})\equiv n^*$, where $n^*$ is the value of the zero-order refractive index that we want to assign to both plasmas. Of course, in order to obtain physically ``sensible'' values of the intensities $I_{0,1}$ and $I_{0,2}$, the two densities cannot have arbitrary values. In Fig. 2 we show a numerical particular example of the outlined procedure. First, we have fixed the laser photon energy equal to $\omega=1.17\;\text{eV}$, the ion charge equal to $Z=46$ (corresponding to palladium) and $n^*=5\times 10^{-2}$. We stress that with these numerical values and in the whole ranges represented in Fig. 2, then $n_{0,j}\approx n_{pl,0,j}=n^*$. Now, the continuous line in Fig. 2 shows the pairs $(N_0,I_0)$ that solve the implicit equation $n_{pl,0}(N_0,I_0)=n^*$. Instead, the dotted line corresponds to the approximated analytical solution [see also Eq. (\ref{T})]
\begin{equation}
\label{I_N}
I_0=I_0(N_0)\approx\frac{(1+n^*)^2}{4}\left\{\left[\frac{e^2N_0}{m_e\omega^2(1-n^{*\,2})}\right]^2-1\right\}\frac{\omega^2}{m^2_e}I_{cr}
\end{equation}
obtained in the limit $m_i\to \infty$ and that, as expected, underestimates the correct intensities when they become of order of $10^{23}\text{-}10^{24}\;\text{W/cm$^2$}$ at which the ion motion cannot be neglected. Following the above procedure, if, for example, $N_{0,1}= 10^{23}\;\text{cm$^{-3}$}$ and $N_{0,2}=2\times 10^{23}\;\text{cm$^{-3}$}$ are the two plasma densities, then we would choose the two intensities as $I_{0,1}=7.2\times 10^{21}\;\text{W/cm$^2$}$ and $I_{0,2}=3.0\times 10^{22}\;\text{W/cm$^2$}$ (see Fig. 2). The necessity to have high plasma densities comes from the fact that the effective plasma frequency $\omega^*_{pl,0}=\omega\sqrt{1-n^{*\,2}}$ must be slightly smaller than the strong laser frequency $\omega$ which lies typically in the optical domain. Moreover, high laser intensities decrease the effective plasma frequency.

By means of the above described procedure, the polarization rotation angle $\Delta\phi(z)$ gives a measure of the vacuum polarization effects and it is given by
\begin{equation}
\label{Delta_phi_f}
\Delta \phi(z)=\frac{16\alpha}{45\pi}\frac{E_2^2-E_1^2}{E^2_{cr}}\frac{(1-n^{*
\,2})^2}{n_{0,1}+n_{0,2}}\omega z\approx\frac{2\alpha}{45\pi}\frac{I_{0,2}-I_{0,1}}{I_{cr}}\frac{(1-n^*)^2}{n^*}\omega z
\end{equation}
where we used the fact that, as we have said, in the previous numerical example $n_{0,j}\approx n_{pl,0,j}=n^*$. By substituting these parameters in Eq. (\ref{Delta_phi_f}) we obtain a rotation angle $\Delta \phi(z_0)=6.8\times 10^{-8}\;\text{rad}$ after a propagation distance of five laser wavelengths in plasma corresponding to $z_0\approx 100\;\text{$\mu$m}$. In order to compare this result with those obtained in vacuum we refer to the light-by-light diffraction process described in Ref. \cite{Di_Piazza_2006}. In fact, as we have observed, one single plane wave in vacuum does not give rise to VPEs. Since we have assumed in our example that the two beams are obtained by splitting an initial laser beam, for a fair comparison we have to use in vacuum the total intensity $I_0=I_{0,1}+I_{0,2}=3.8\times 10^{22}\;\text{W/cm$^2$}$ and we obtain $\Delta \phi_{\text{vac}}=3.8\times 10^{-9}\;\text{rad}$ which is more than one order of magnitude less than $\Delta \phi(z_0)$. This result in vacuum can be easily derived by looking at the quantity indicated as $\psi$ in Fig. 2 in Ref. \cite{Di_Piazza_2006}. It shows the polarization rotation angle that an x-ray probe undergoes by interacting with a strong optical standing wave as a function of the distance between the photon polarimeter and the lasers collision region. In the figure the strong optical wave intensity is $10^{23}\;\text{W/cm$^2$}$ and, since the dependence on the polarimeter distance is weak and since the rotation angle is proportional to the strong optical laser intensity, the desired rotation polarization angle can be obtained by multiplying the average value of the rotation angle in the figure, which is approximately $10^{-8}\;\text{rad}$, times $I_0[\text{W/cm$^2$}]/10^{23}$. 

We mention here that the tuning of the intensities $I_{0,j}$ to eliminate the pure plasma contribution to $\Delta \phi(z)$ can be avoided by performing different experiments with different laser peak intensities $I_0$. In this case, one could reveal the presence of VPEs by fitting the resulting experimental curve $\Delta \phi(z)=\Delta \phi(z,I_0)$ by using Eq. (\ref{Delta_phi}) with and without the inclusion of VPEs.

From the previous analysis we can conclude that the presence of a plasma can give rise to a large enhancement of the VPEs. This result is very relevant also because, as it is clear from the above example, in plasma there is a significant disadvantage (that has been accounted for) with respect to vacuum. In fact, in vacuum one can use high-frequency, i.e. x-ray, probes for which the phase difference is proportionally larger than that for optical probes. Instead in plasma, since the probe field (that here is the same as the strong field) must have a frequency close to the plasma frequency one is forced to use at most optical probes that already require high electron densities of order of $10^{23}\;\text{cm$^{-3}$}$.

Despite the above results, we are aware of the practical difficulties that are behind the experimental realization of the previous setup. The two intensities $I_{0,1}$ and $I_{0,2}$ must be carefully adjusted in order to eliminate the spurious zero-order effects in $\Delta \phi(z)$ due to the presence of the plasma. Also, since the strong waves have frequencies close to the effective plasma frequency different kinds of plasma instabilities can arise like, for example, the so-called parametric resonance (see Ref. \cite{Borovski_b_2003} and the references therein). Finally, other difficulties have to be mentioned that are connected with the basic assumptions of our model. For example, the monochromatic-wave approximation has allowed us to isolate, at least theoretically, the VPEs and to obtain an analytical estimate of them. Nevertheless, if we considered a more realistic pulsed wave with a transverse profile then the treatment would be much more complex and many other effects should be taken into account like laser self-focusing, channeling, filamentation and so on (see Ref. \cite{Umstadter_2003} and the references therein). A further remark is in order because some effects can hinder the laser penetration into the plasma. We mention here the parametric instabilities and the plasma density profile steepening. On the one hand, the excitation time of the Raman instability is extremely short, about half of the laser period \cite{Mora}, excluding the possibility to avoid it by using short laser pulses. Nevertheless, the parametric instabilities saturate due to nonlinear mechanisms before the laser beam will be depleted \cite{Rousseaux}. Moreover, the transmission of the laser wave through the plasma improves at laser intensities exceeding $10^{20}$W/cm$^2$ \cite{Adam}, and almost half of the laser pulse can be transmitted through the plasma on the length of about $200$ wavelengths. On the other hand, the profile steepening of the electron and ion densities due to the laser ponderomotive force can play a more important role. Because of this effect the laser intensity threshold for induced transparency is significantly raised at a given plasma density, or the density threshold is decreased at a fixed laser intensity \cite{Lisak}. As we have observed, we have chosen high plasma densities of order of $10^{23}\;\text{cm$^{-3}$}$ to fulfill the condition of the near overdense plasma $\omega\gtrsim\omega_{pl,0}$ with $\omega$ in the optical region. Instead, we could exploit the density profile steepening to consider smaller initial plasma densities. In this case one could expect the effect of enhancement of VPEs to be present at the new resulting threshold of the plasma transparency but only in the steepening region. This is because the enhancement effect arises when the effective refractive index of the plasma tends to zero. In order to discuss quantitatively all the above effects a detailed analysis in more realistic conditions is required. This would include, for example, particle-in-cell simulations and it is out of the scope of the present treatment that is meant to be essentially qualitative.

One final comment concerns the practical difficulty in measuring rotation polarization angles $\Delta \phi$ with an accuracy of the order of vacuum polarization corrections, that is $10^{-8}\;\text{rad}$ in our numerical example. This problem is, of course, present also in the pure vacuum case independently of plasma. Nevertheless, very sensitive polarimeters are available today in the optical regime that are able in principle to measure polarization rotation angles with an accuracy of order of $10^{-8}\;\text{rad}$ \cite{Muroo_2000}. Also, the number of photons delivered in one pulse by an optical laser with intensity of order of $10^{22}\;\text{W/cm$^2$}$ is much larger than the number of photons $n\sim  10^{16}$ required from statistical considerations to measure polarization rotation angles with an accuracy of order of $10^{-8}\;\text{rad}$.

All the above analysis has been carried out by neglecting the collisional effects. This can be a too crude approximation mostly because we are interested in the physical regime where the wave frequency approaches the plasma frequency. We will include the effects of electron-ion collisions in the next Section and, actually, it will appear that the previous experimental setup has to be modified.
%
%
\section{Inclusion of collisional effects}
\label{Collisions}
In the previous Section we have considered the propagation of a strong wave through a collisionless plasma in order to point out the role of the VPEs and to show how they are enhanced with respect to pure vacuum and how they can be detected experimentally. The enhancement is significant when the laser frequency is slightly larger than the plasma frequency. As we have already observed, in this parameter region the inclusion of collisional effects is conceptually relevant. In this Section we want to investigate this case that will turn out technically more complex than the collisionless case already considered. Nevertheless, we want to point out here that the final conclusions about the enhancement and the measurability of VPEs will not be changed by the inclusion of the collisional effects. Moreover, the analytical implicit solution we will find in the following Paragraph [see, in particular, Eqs. (\ref{E_fin_coll})-(\ref{v}), (\ref{h_p}) and (\ref{g_appr})] is of interest by itself independently of VPEs. In fact, it gives an accurate description of the wave propagation through the plasma in the limit of small collision frequencies with respect to the laser frequency.
\subsection{Analysis of the VPEs for the strong wave propagation through a collisional plasma}
Starting again from our initial Eqs. (\ref{Div_D})-(\ref{Eq_mot_p}) we can assume here that all physical quantities depend only on the adimensional variables $\varphi=kz$ and $\eta=\omega t$. Also, we are interested in the propagation of a \emph{transverse} electromagnetic wave. In this case it is possible, as before, to look for a solution of the equations such that the electron and ion charge densities are equal to each other and constant and uniform, and the two fluids velocities $\mathbf{v}_{\lambda}(\varphi,\eta)$ are transverse:
\begin{gather}
N_e(\varphi,\eta)=ZN_i(\varphi,\eta)\equiv N_0,\\
\mathbf{k}\cdot \mathbf{v}_{\lambda}(\varphi,\eta)=0
\end{gather}
where $\mathbf{k}=k\hat{\mathbf{z}}$ is the wave vector. Since the fluids velocities are transverse and all the quantities depend only on time and on the space coordinate $z$, it follows that in Eqs. (\ref{Eq_mot_e}) and (\ref{Eq_mot_p})
\begin{equation}
(\mathbf{v}_{\lambda}(\varphi,\eta)\cdot\boldsymbol{\partial})\mathbf{p}_{\lambda}(\varphi,\eta)=\mathbf{0}.
\end{equation}
Also, by multiplying Eqs. (\ref{Eq_mot_e}) and (\ref{Eq_mot_p}) times $\mathbf{k}$ we conclude that $\mathbf{v}_{\lambda}(\varphi,\eta)\parallel\mathbf{B}(\varphi,\eta)$. Now, in order to guarantee the total momentum conservation during the collisions, the collision frequencies $\nu_{ei}$ and $\nu_{ie}$ must satisfy the condition $N_e\nu_{ei}=N_i\nu_{ie}$, that is $Z\nu_{ei}=\nu_{ie}$. From this relation and from Eqs. (\ref{Eq_mot_e}) and (\ref{Eq_mot_p}) it can be seen that a solution exists such that
\begin{equation}
\mathbf{p}_e(\varphi,\eta)=-\frac{\mathbf{p}_i(\varphi,\eta)}{Z}\equiv\mathbf{p}(\varphi,\eta).
\end{equation}
Finally, by introducing the vector potential $\mathbf{A}(\varphi,\eta)$ such that
\begin{align}
\label{E_A}
\mathbf{E}(\varphi,\eta)&=-\omega\partial_{\eta}\mathbf{A}(\varphi,\eta),\\
\label{B_A}
\mathbf{B}(\varphi,\eta)&=\mathbf{k}\times\partial_{\varphi}\mathbf{A}(\varphi,\eta),
\end{align}
the initial set of equations (\ref{Div_D})-(\ref{Eq_mot_p}) reduces to the following coupled nonlinear equations for the quantities $\mathbf{A}(\varphi,\eta)$ and $\mathbf{p}(\varphi,\eta)$ (in particular, Eq. (\ref{Eq_A}) comes from Eq. (\ref{Rot_H}), and Eq. (\ref{Eq_p}) from Eq. (\ref{Eq_mot_e})):
\begin{align}
\label{Eq_A}
\mathbf{k}\times\partial_{\varphi}\mathbf{B}-\omega\partial_{\eta}\mathbf{E}&=-eN_0\left(\frac{1}{m_e\gamma_e}+\frac{Z}{m_i\gamma_i}\right)\mathbf{p}+\mathbf{J}_{\text{vac}}(\mathbf{A}),\\
\label{Eq_p}
\omega\partial_{\eta}\mathbf{p}&=e\omega\partial_{\eta}\mathbf{A}-\nu\mathbf{p}
\end{align}
with $\gamma_{\lambda}=\gamma_{\lambda}(\varphi,\eta)=\sqrt{1+Z_{\lambda}^2p^2(\varphi,\eta)/m^2_{\lambda}}$ and
\begin{equation}
\label{nu}
\nu\equiv\nu_{ei}\frac{m_i+Zm_e}{m_i+m_e}.
\end{equation}
For notational simplicity we have implied that the current $\mathbf{J}_{\text{vac}}(\varphi,\eta)$ has to be expressed in terms of the vector potential $\mathbf{A}(\varphi,\eta)$ [see Eqs. (\ref{J_v}), (\ref{E_A}) and (\ref{B_A})].

In order to eliminate the dependence on the variable $\eta$, we look for a solution of the previous equations of the form
\begin{align}
\label{A}
\mathbf{A}(\varphi,\eta)&=\mathbf{A}_c(\varphi)\cos\eta+\mathbf{A}_s(\varphi)\sin\eta,\\
\label{p}
\mathbf{p}(\varphi,\eta)&=\mathbf{p}_c(\varphi)\cos\eta+\mathbf{p}_s(\varphi)\sin\eta
\end{align}
with the conditions
\begin{align}
\label{Cond_p_1}
\mathbf{p}_c(\varphi)\cdot\mathbf{p}_s(\varphi)&=0,\\
\label{Cond_p_2}
p_c^2(\varphi)=p_s^2(\varphi)&=p^2(\varphi).
\end{align}
It is easy to see that in this way the differential equation (\ref{Eq_p}) becomes algebraic and its solution is
\begin{align}
\mathbf{p}_c(\varphi)&=\frac{e\omega}{\omega^2+\nu^2}[\omega\mathbf{A}_c(\varphi)+\nu\mathbf{A}_s(\varphi)],\\
\mathbf{p}_s(\varphi)&=\frac{e\omega}{\omega^2+\nu^2}[\omega\mathbf{A}_s(\varphi)-\nu\mathbf{A}_c(\varphi)].
\end{align}
The momentum components expressed in this form satisfy the conditions (\ref{Cond_p_1}) and (\ref{Cond_p_2}) if the analogous following conditions are satisfied by the vector potential components $\mathbf{A}_c(\varphi)$ and $\mathbf{A}_s(\varphi)$:
\begin{align}
\label{Cond_A_1}
\mathbf{A}_c(\varphi)\cdot\mathbf{A}_s(\varphi)&=0,\\
\label{Cond_A_2}
A_c^2(\varphi)=A_s^2(\varphi)&=A^2(\varphi).
\end{align}
At this point we have to solve only Eq. (\ref{Eq_A}) for the vector potential components. In fact, by using the above equations we can express also the relativistic Lorentz factors only in terms of the vector potential amplitude $A(\varphi)$ as
$\gamma_{\lambda}(\varphi)=\sqrt{1+Z_{\lambda}^2e^2\omega^2A^2(\varphi)/m^2_{\lambda}(\omega^2+\nu^2)}$. Now, the conditions (\ref{Cond_A_1}) and (\ref{Cond_A_2}) suggest to look for the vectors $\mathbf{A}_c(\varphi)$ and $\mathbf{A}_s(\varphi)$ of the general form
\begin{align}
\label{A_c}
\mathbf{A}_c(\varphi)&=Ae^{-h(\varphi)}[\hat{\mathbf{x}}\cos g(\varphi)+\hat{\mathbf{y}}\sin g(\varphi)],\\
\label{A_s}
\mathbf{A}_s(\varphi)&=Ae^{-h(\varphi)}[\hat{\mathbf{x}}\sin g(\varphi)-\hat{\mathbf{y}}\cos g(\varphi)]
\end{align}
where the amplitude $A=-E/\omega$ is a negative quantity. With these expressions the total vector potential (\ref{A}), the electron momentum (\ref{p}), the electric field (\ref{E_A}) and the magnetic field (\ref{B_A}) are given by
\begin{align}
\label{A_fin_coll}
\mathbf{A}(\varphi,\eta)&=Ae^{-h(\varphi)}\mathbf{u}(\varphi,\eta),\\
\label{p_fin_coll}
\mathbf{p}(\varphi,\eta)&=-\frac{eE}{\omega}e^{-h(\varphi)}\left[\frac{\omega^2}{\omega^2+\nu^2}\mathbf{u}(\varphi,\eta)+\frac{\omega\nu}{\omega^2+\nu^2}\mathbf{v}(\varphi,\eta)\right]\\
\label{E_fin_coll}
\mathbf{E}(\varphi,\eta)&=Ee^{-h(\varphi)}\mathbf{v}(\varphi,\eta),\\
\label{B_fin_coll}
\mathbf{B}(\varphi,\eta)&=nEe^{-h(\varphi)}[-h'(\varphi)\mathbf{v}(\varphi,\eta)+g'(\varphi)\mathbf{u}(\varphi,\eta)]
\end{align}
with $n=k/\omega$. In the above expressions the prime denotes the derivative with respect to $\varphi$ and the two perpendicular unit vectors
\begin{align}
\label{u}
\mathbf{u}(\varphi,\eta)&\equiv\hat{\mathbf{x}}\cos (g(\varphi)-\eta)+\hat{\mathbf{y}}\sin (g(\varphi)-\eta),\\
\label{v}
\mathbf{v}(\varphi,\eta)&\equiv\hat{\mathbf{x}}\sin (g(\varphi)-\eta)-\hat{\mathbf{y}}\cos (g(\varphi)-\eta)
\end{align}
such that
\begin{align}
\partial_{\varphi}\mathbf{u}(\varphi,\eta)&=-g'(\varphi)\mathbf{v}(\varphi,\eta), & \partial_{\eta}\mathbf{u}(\varphi,\eta)&=\mathbf{v}(\varphi,\eta), & \mathbf{k}\times\mathbf{u}(\varphi,\eta)&=-n\omega\mathbf{v}(\varphi,\eta)\\
\partial_{\varphi}\mathbf{v}(\varphi,\eta)&=g'(\varphi)\mathbf{u}(\varphi,\eta), &
\partial_{\eta}\mathbf{v}(\varphi,\eta)&=-\mathbf{u}(\varphi,\eta), & \mathbf{k}\times\mathbf{v}(\varphi,\eta)&=n\omega\mathbf{u}(\varphi,\eta)
\end{align}
have been introduced. By using Eqs. (\ref{E_fin_coll}) and (\ref{B_fin_coll}) we obtain that the two quantities
\begin{align}
\label{Inv_1}
E^2(\varphi,\eta)-B^2(\varphi,\eta)&=E^2e^{-2h(\varphi)}\{1-n^2[h^{\prime\,2}(\varphi)+g^{\prime\,2}(\varphi)]\},\\
\label{Inv_2}
\mathbf{E}(\varphi,\eta)\cdot\mathbf{B}(\varphi,\eta)&=nE^2e^{-2h(\varphi)}h'(\varphi)
\end{align}
needed to calculate the vector $\mathbf{J}_{\text{vac}}(\varphi,\eta)$ depend only on $\varphi$. This useful property is related to the fact that we are considering a circularly polarized wave. It also allows, by using the properties of the unit vectors (\ref{u}) and (\ref{v}), to easily eliminate the dependence on $\eta$ in Eq. (\ref{Eq_A}). For the sake of clarity we indicate here the expression of the current $\mathbf{J}_{\text{vac}}(\varphi,\eta)$ in terms of the functions $h(\varphi)$ and $g(\varphi)$ and of the unit vectors $\mathbf{u}(\varphi,\eta)$ and $\mathbf{v}(\varphi,\eta)$. Starting from Eq. (\ref{J_v}) and by using the expressions (\ref{Inv_1}) and (\ref{Inv_2}) of the electromagnetic invariants, we obtain (for notational simplicity we omit the dependence of the various quantities on $\varphi$ and $\eta$)
\begin{equation}
\begin{split}
\mathbf{J}_{\text{vac}}&=-\frac{4\alpha^2}{45m_e^4}\left\{\mathbf{k}\times\partial_{\varphi}\left[2(E^2-B^2)\mathbf{B}-7(\mathbf{E}\cdot\mathbf{B})\mathbf{E}\right]-\omega\partial_{\eta}\left[2(E^2-B^2)\mathbf{E}+7(\mathbf{E}\cdot\mathbf{B})\mathbf{B}\right]\right\}\\
&=\frac{4\alpha^2}{45m_e^4}\omega Ee^{-h}\left\{2n^2(E^2-B^2)[(h''-h^{\prime 2}+g^{\prime 2})\mathbf{u}+(g''-2h'g')\mathbf{v}]\right.\\
&\left.\qquad\qquad\qquad\quad+2n^2(h'\mathbf{u}+g'\mathbf{v})\partial_{\varphi}(E^2-B^2)+[7n\partial_{\varphi}(\mathbf{E}\cdot\mathbf{B})-2(E^2-B^2)]\mathbf{u}\right\}\\
&=\omega Ee^{-h}\frac{\alpha}{45\pi}\frac{E^2}{E^2_{cr}}e^{-2h}\big\{2n^2[1-n^2(h^{\prime\,2}+g^{\prime\,2})][(h''-h^{\prime 2}+g^{\prime 2})\mathbf{u}+(g''-2h'g')\mathbf{v}]\\
&\qquad\qquad\qquad\qquad\qquad-4n^2(h'\mathbf{u}+g'\mathbf{v})\big[h'[1-n^2(h^{\prime\,2}+g^{\prime\,2})]+n^2(h'h''+g'g'')\big]\\
&\qquad\qquad\qquad\qquad\qquad+\big[7n^2(h''-2h^{\prime\,2})-2[1-n^2(h^{\prime\,2}+g^{\prime\,2})]\big]\mathbf{u}\big\}\\
\end{split}
\end{equation}
By substituting this expression of $\mathbf{J}_{\text{vac}}(\varphi,\eta)$, and the expressions (\ref{p_fin_coll}), (\ref{E_fin_coll}) and (\ref{B_fin_coll}) of $\mathbf{p}(\varphi,\eta)$, $\mathbf{E}(\varphi,\eta)$ and $\mathbf{B}(\varphi,\eta)$ into Eq. (\ref{Eq_A}) and by separating the terms proportional to $\mathbf{u}(\varphi,\eta)$ from those proportional to $\mathbf{v}(\varphi,\eta)$, we obtain the following final equations for the unknown functions $h(\varphi)$ and $g(\varphi)$:
\begin{align}
\label{f}
\begin{split}
&\left\{1+\frac{2\alpha}{45\pi}\frac{E^2}{E^2_{cr}}e^{-2h}[1-n^2(h^{\prime\,2}+g^{\prime\,2})]\right\}[1+n^2(-h''+h^{\prime 2}-g^{\prime 2})]=\frac{1}{\omega^2}\left(\frac{\omega^2_{pl,e}}{\gamma_e(h)}+\frac{\omega^2_{pl,i}}{\gamma_i(h)}\right)\frac{\omega^2}{\omega^2+\nu^2}-\\
&\qquad-\frac{\alpha}{45\pi}\frac{E^2}{E^2_{cr}}e^{-2h}n^2\{4h'[h'[1-n^2(h^{\prime\,2}+g^{\prime\,2})]+n^2(h'h''+g'g'')]-7(h''-2h^{\prime\,2})\},
\end{split}\\
\label{g}
\begin{split}
&\left\{1+\frac{2\alpha}{45\pi}\frac{E^2}{E^2_{cr}}e^{-2h}[1-n^2(h^{\prime\,2}+g^{\prime\,2})]\right\}
(-2h'g'+g'')=-\frac{1}{n^2\omega^2}\left(\frac{\omega^2_{pl,e}}{\gamma_e(h)}+\frac{\omega^2_{pl,i}}{\gamma_i(h)}\right)\frac{\omega\nu}{\omega^2+\nu^2}+\\
&\qquad+\frac{4\alpha}{45\pi}\frac{E^2}{E^2_{cr}}e^{-2h}g'\{h'[1-n^2(h^{\prime\,2}+g^{\prime\,2})]+n^2(h'h''+g'g'')\}
\end{split}
\end{align}
where
\begin{equation}
\label{gamma}
\gamma_{\lambda}(\varphi)=\sqrt{1+\frac{Z_{\lambda}^2e^2}{m^2_{\lambda}}\frac{E^2}{\omega^2+\nu^2}e^{-2h(\varphi)}}
\end{equation}
are the relativistic Lorentz factors due to the particles motion in the wave field.

In conclusion, we have reduced the initial equations (\ref{Div_D})-(\ref{Eq_mot_p}) to the above Eqs. (\ref{f}) and (\ref{g}). These equations cannot be solved exactly if $\nu\neq 0$ but we can exploit the fact that in the presence of a strong wave the parameter $\delta_{\nu}\equiv\nu/\omega$ can be considered very small. To give a quantitative estimate we use the usual relativistic transformation formulas \cite{Landau_b_2_1975} and we can write the quantity $\nu$ in Eq. (\ref{nu}) as
\begin{equation}
\nu=\frac{m_i+Zm_e}{m_i+m_e}\nu_{ei}=\frac{m_i+Zm_e}{m_i+m_e}\left(1+\frac{Zp_0^2}{m_em_i\gamma_{0,e}\gamma_{0,i}}\right)\nu^*_{ei}
\end{equation}
with $\nu^*_{ei}$ being the effective electron-ion collision frequency in the reference system where the ions are at rest. In our units $\nu^*_{ei}$ is given by \cite{Hora_b_1991}
\begin{equation}
\nu^*_{ei}=\frac{\pi^{3/2}ZN_0}{4\sqrt{2m_e\gamma^*_{0,e}}}\frac{e^4}{(4\pi)^2}\frac{1}{m_e^{3/2}(\gamma^*_{0,e}-1)^{3/2}}\log\left[\frac{12\pi}{Ze^3}\left(\frac{m_e^3(\gamma^*_{0,e}-1)^3}{N_0}\right)^{1/2}\right]
\end{equation}
where $\gamma^*_{0,e}=(1-v_{0,e}^{*\,2})^{-1/2}$ is the electron relativistic Lorentz factor in the same reference system with
\begin{equation}
\mathbf{v}^*_{0,e}=\frac{m_i\gamma_{0,i}+Zm_e\gamma_{0,e}}{m_em_i\gamma_{0,e}\gamma_{0,i}+Zp_0^2}\mathbf{p}_0.
\end{equation}
The index ``0'' in the dynamical quantities in the above expressions means that they are calculated by solving Eq. (\ref{f}) and (\ref{g}) with $\nu=0$ and they coincide with those introduced in Sec. \ref{Results}\cite{Footnote_2}. By using the results already obtained there, it is easy to show that the condition $\delta_{\nu}\ll 1$ is fulfilled for electric field amplitudes $E$ such that
\begin{equation}
\frac{E^2}{E^2_{cr}}\gg \frac{ZN_0}{m_e^3}\frac{\omega}{m_e} 
\end{equation}
where for simplicity we have assumed $m_i\to\infty$ and we have neglected the logarithm correction to $\nu$. We note that the right hand side of this strong inequality is usually a quantity several orders of magnitude smaller than unity. For example, if $Z=10$, $N_0=10^{23}\;\text{cm$^{-3}$}$ and $\omega=1.17\;\text{eV}$ then $ZN_0\omega/m_e^4\sim 10^{-14}$. 

Now, from the zero-order solution given in Sect. \ref{Results} we realize that in the limit $\delta_{\nu}\ll 1$ the functions $h(\varphi)$ and $g'(\varphi)$ are slowly varying because $h^{(0)}(\varphi)=0$ and $g^{(0)}(\varphi)=\varphi$. This means that generally speaking $h'(\varphi),g''(\varphi)\sim \delta_{\nu}$ and $h''(\varphi)\sim\delta_{\nu}^2$. Moreover, as we have mentioned before, we have taken into account the VPEs up to first order in $\delta_{\text{vac}}$ already in the initial equations (\ref{Div_D})-(\ref{Eq_mot_p}). For these reasons we will limit ourselves to find a solution of Eqs. (\ref{f}) and (\ref{g}) up to first order in the two limits $\delta_{\nu}\ll 1$ and $\delta_{\text{vac}}\ll 1$. In these approximations Eqs. (\ref{f}) and (\ref{g}) assume the simplified form
\begin{align}
\label{f_nu_small}
1-n_0^2g^{\prime 2}&=\frac{1}{\omega^2}\left(\frac{\omega^2_{pl,e}}{\gamma_e(h)}+\frac{\omega^2_{pl,i}}{\gamma_i(h)}\right)\left[1-\frac{2\alpha}{45\pi}\frac{E^2}{E^2_{cr}}\frac{e^{-2h}}{\omega^2}\left(\frac{\omega^2_{pl,e}}{\gamma_e(h)}+\frac{\omega^2_{pl,i}}{\gamma_i(h)}\right)\right],\\
\label{g_nu_small}
g''-2h'g'&=-\frac{1}{n_0^2\omega^2}\left(\frac{\omega^2_{pl,e}}{\gamma_e(h)}+\frac{\omega^2_{pl,i}}{\gamma_i(h)}\right)\frac{\nu}{\omega}.
\end{align}
We observe that in these equations the two relativistic Lorentz factors are given by $\gamma_{\lambda}(h(\varphi))=\sqrt{1+Z_{\lambda}^2e^2E^2e^{-2h(\varphi)}/\omega^2m^2_{\lambda}}$ that are different from Eq. (\ref{gamma}): for notational simplicity we have used the same symbol because we will not use anymore the expressions in Eq. (\ref{gamma}). The substitution $n=n_0$ we have tacitly made in Eqs. (\ref{f_nu_small}) and (\ref{g_nu_small}) results automatically by imposing the initial conditions $h(0)=0$ on the wave amplitude and $g(0)=0$ and $g'(0)=1$ on the wave phase [see Eq. (\ref{f_nu_small})]. In turn, these conditions ensure that the solution of Eqs. (\ref{f_nu_small}) and (\ref{g_nu_small}) reduce to the Akhiezer-Polovin solution plus VPEs if $\nu=0$. In this respect, we observe that since $h(0)=0$ and $h'(\varphi)\sim \delta_{\nu}$ then also $h(\varphi)\sim \delta_{\nu}$. Nevertheless, we have treated more exactly the terms in $h(\varphi)$ in order to have a solution valid also for large values of $\varphi$. Now, on one hand, we obtain from Eq. (\ref{g_nu_small}) that the function $g'(\varphi)$ can be expressed in terms of the function $h(\varphi)$ as
\begin{equation}
\label{g_p_1}
g'(\varphi)=e^{2h(\varphi)}\left[1-\frac{\nu}{\omega}\int_0^{\varphi}d\varphi'e^{-2h(\varphi')}\frac{1}{n_0^2\omega^2}\left(\frac{\omega^2_{pl,e}}{\gamma_e(h(\varphi'))}+\frac{\omega^2_{pl,i}}{\gamma_i(h(\varphi'))}\right)\right].
\end{equation}
On the other hand, if we introduce the quantity
\begin{equation}
\label{n_h}
n(h(\varphi))=\sqrt{1-\frac{1}{\omega^2}\left(\frac{\omega^2_{pl,e}}{\gamma_{0,e}(h(\varphi))}+\frac{\omega^2_{pl,i}}{\gamma_{0,i}(h(\varphi))}\right)+\frac{2\alpha}{45\pi}\frac{E^2}{E^2_{cr}}\frac{e^{-2h(\varphi)}}{\omega^4}\left(\frac{\omega^2_{pl,e}}{\gamma_{0,e}(h(\varphi))}+\frac{\omega^2_{pl,i}}{\gamma_{0,i}(h(\varphi))}\right)^2},
\end{equation}
then Eq. (\ref{f_nu_small}) becomes
\begin{equation}
\label{g_p_2}
g'(\varphi)=\frac{n(h(\varphi))}{n_0}.
\end{equation}
Finally, by combining Eqs. (\ref{g_p_1}) and (\ref{g_p_2}) we obtain the following differential equation for the function $h(\varphi)$:
\begin{equation}
\label{h_p}
h'=\frac{\nu}{\omega}\frac{n(h)}{n_0}\frac{\frac{1}{\omega^2}\left(\frac{\omega^2_{pl,e}}{\gamma_e(h)}+\frac{\omega^2_{pl,i}}{\gamma_i(h)}\right)}{2n^2(h)+\frac{1}{2\omega^2}\left(\frac{\omega^2_{pl,e}}{\gamma_e(h)}\frac{e^2E^2e^{-2h}}{\omega^2m_e^2+e^2E^2e^{-2h}}+\frac{\omega^2_{pl,i}}{\gamma_i(h)}\frac{Z^2e^2E^2e^{-2h}}{\omega^2m_i^2+Z^2e^2E^2e^{-2h}}\right)}.
\end{equation}
This equation can in principle be integrated and, by substituting the solution into Eq. (\ref{g_p_2}) we obtain
\begin{equation}
\label{g_appr}
g(\varphi)=\varphi-\int_0^{\varphi}d\varphi'\left(1-\frac{n(h(\varphi'))}{n_0}\right).
\end{equation}
In conclusion, due to the electron-ion collisions both the amplitude and the phase of the wave are modified. As expected, the amplitude decreases during the propagation ($h(0)=0$ and $h'(\varphi)>0$). Also, by neglecting for simplicity the vacuum corrections in $n(h(\varphi))$ we can say that if the wave frequency is larger than the non-relativistic plasma frequency $\sqrt{\omega^2_{pl,e}+\omega^2_{pl,i}}$ then $n(h(\varphi))>0$ for every $\varphi$. This means that $h'(\varphi)>0$ for every $\varphi$ and then that the wave amplitude goes to zero for very large $\varphi$. Instead, in the complementary case where $\omega<\sqrt{\omega^2_{pl,e}+\omega^2_{pl,i}}$, the function $h(\varphi)$ tends asymptotically to the value $h^*$ such that $n(h^*)=0$ that physically corresponds to the impossibility of further wave propagation. 

In order to determine the physical conditions in which the above solution holds, we observe that in passing from Eqs. (\ref{f}) and (\ref{g}) to Eqs. (\ref{f_nu_small}) and (\ref{g_nu_small}) we have used the following strong inequalities valid in the regime $\omega\gtrsim\omega_{pl,0}$:
\begin{align}
h''(\varphi),h^{\prime\, 2}(\varphi)&\ll g^{\prime\, 2}(\varphi),\\
\delta_{\nu},\delta_{\text{vac}}&\ll 1,\\
\delta_{\text{vac}}n_0^2g'(\varphi)h'(\varphi),\delta_{\text{vac}}n_0^4g^{\prime\, 2}(\varphi)(\varphi)g''(\varphi)&\ll \delta_{\nu}.
\end{align}
Now, Eq. (\ref{g_p_2}) implies that $g'(\varphi)\lesssim 1$. Also, while $h'(\varphi)\sim \delta_{\nu}g'(\varphi)$, the second derivatives $h''(\varphi)$ and $g''(\varphi)$ can be estimated from Eqs. (\ref{h_p}) and (\ref{g_p_2}) as $h''(\varphi)\sim \delta_{\nu}^2/n_0^2$ and $g''(\varphi)\sim \delta_{\nu}/n_0^2$. By using these estimations it can be easily shown that our approximated solution holds if only the two following conditions are fulfilled:
\begin{align}
\label{cond_provv_1}
\delta^2_{\nu}&\ll n^2(h(\varphi)),\\
\label{cond_provv_2}
\delta_{\text{vac}}&\ll 1.
\end{align}
The first of the above conditions implies that at very large $\varphi$ our solution is no more valid because, as we have said, for wave frequencies close to $\omega_{pl,0}$ then $\lim_{\varphi\to\infty}n(h(\varphi))=0$. Nevertheless, we can qualitatively say from Eq. (\ref{h_p}) that if we choose the initial parameters such that $\delta^2_{\nu}\ll n^2_0$ then the condition (\ref{cond_provv_1}) is satisfied up to values of $\varphi$ of order of $1/\delta^2_{\nu}\gg 1$. For example, we have seen numerically that at wave intensities of order of $10^{22}\;\text{W/cm$^2$}$ then $\delta_{\nu}\sim 10^{-10}$ for an optical frequency $\omega$ and, in conclusion, the condition in Eq. (\ref{cond_provv_1}) is in practice not restrictive at all. Finally, we have also to impose that $\delta^2_{\text{vac}}\ll \delta_{\nu}$ to guarantee that the second-order corrections with respect to VPEs are consistently negligible. This condition is stronger than the one in Eq. (\ref{cond_provv_2}) then our final validity-conditions are
\begin{align}
\label{cond_1}
\delta^2_{\nu}&\ll n^2(h(\varphi)),\\
\label{cond_2}
\delta^2_{\text{vac}}&\ll \delta_{\nu}.
\end{align}
It is worth pointing out here that, independently on VPEs, the above implicit solution (\ref{h_p})-(\ref{g_appr}) holds under very general conditions and it can represent by itself a useful tool to study analytically the propagation of a strong wave through a plasma by including the effects of the electron-ion collisions. 

In order to obtain a more transparent solution we observe that if the function $h(\varphi)$ itself is very small then we can set $h(\varphi)=0$ in the right hand side of Eq. (\ref{h_p}) and $h(\varphi)\approx h'(0)\varphi$. In these approximations we obtain the simplified explicit solution [see also Eq. (\ref{g_appr})]
\begin{align}
\label{h_appr_appr}
h(\varphi)&\approx\frac{\nu}{\omega}\frac{1}{\omega^2}\frac{\frac{\omega^2_{pl,e}}{\gamma_{0,e}}+\frac{\omega^2_{pl,i}}{\gamma_{0,i}}}{2n_0^2+\frac{1}{2\omega^2}\left(\frac{\omega^2_{pl,e}}{\gamma_{0,e}}\frac{e^2E^2}{\omega^2m_e^2+e^2E^2}+\frac{\omega^2_{pl,i}}{\gamma_{0,i}}\frac{Z^2e^2E^2}{\omega m_i^2+Z^2e^2E^2}\right)}\varphi,\\
\label{g_appr_appr}
g(\varphi)&\approx\varphi-\frac{1}{2}\frac{\nu}{\omega}\frac{1}{n_0^2\omega^2}\left(\frac{\omega^2_{pl,e}}{\gamma_{0,e}}+\frac{\omega^2_{pl,i}}{\gamma_{0,i}}\right)\frac{\frac{1}{\omega^2}\left(\frac{\omega^2_{pl,e}}{\gamma_{0,e}}\frac{e^2E^2}{\omega^2m_e^2+e^2E^2}+\frac{\omega^2_{pl,i}}{\gamma_{0,i}}\frac{Z^2e^2E^2}{\omega m_i^2+Z^2e^2E^2}\right)}{2n_0^2+\frac{1}{2\omega^2}\left(\frac{\omega^2_{pl,e}}{\gamma_{0,e}}\frac{e^2E^2}{\omega^2m_e^2+e^2E^2}+\frac{\omega^2_{pl,i}}{\gamma_{0,i}}\frac{Z^2e^2E^2}{\omega^2m_i^2+Z^2e^2E^2}\right)}\frac{\varphi^2}{2}.
\end{align}
While the approximated expression (\ref{h_appr_appr}) of the function $h(\varphi)$ holds if $\varphi\delta_{\nu}\ll 1$ instead Eq. (\ref{g_appr_appr}) requires that the more restrictive condition $\varphi\delta_{\nu}\ll n^2_0$ is fulfilled. In the following we will always refer to this approximated solution. To the sake of clarity we show in Fig. 3 a typical behaviour of the amplitude of the electric field (\ref{E_fin_coll}) divided by $E$ as a function of the adimensional variable $\varphi$ with $\omega=1.1\sqrt{\omega^2_{pl,e}+\omega^2_{pl,i}}$. The Figure clearly shows the exponential damping of the laser wave. The vertical line corresponds to $\varphi_0\approx 1/10\delta_{\nu}$ and it is clear that, as predicted theoretically, the approximated solution (\ref{h_appr_appr}) works well for $\varphi\lesssim\varphi_0$.
%
%
\subsection{Modification of the experimental setup due to the inclusion of collisional effects}
\label{Setup_Rev}
Now that we have included the collisional effects in the analysis of the strong laser wave evolution in plasma [see Eqs. (\ref{h_appr_appr}) and (\ref{g_appr_appr})], we want to discuss which modifications are needed in the possible experimental setup for the detection of VPEs in plasma discussed in Sec. \ref{Setup}. In the approximations discussed the electric field of the traveling wave is given by [see Eqs. (\ref{E_fin_coll}) and (\ref{v})]
\begin{equation}
\label{E_fin_fin}
\begin{split}
\mathbf{E}(z,t)&=Ee^{-h'(0)\omega n_0z}\left[\hat{\mathbf{x}}\sin \left(\omega(n_0z-t)+g''(0)(\omega n_0z)^2/2\right)\right.\\
&\left.\qquad\qquad\qquad\quad-\hat{\mathbf{y}}\cos \left(\omega(n_0z-t)+g''(0)(\omega n_0z)^2/2\right)\right].
\end{split}
\end{equation}
The refractive index $n_0$ is given in Eq. (\ref{n_0}) and the expressions of the derivatives $h'(0)$ and $g''(0)$ can be easily deduced from Eqs. (\ref{h_appr_appr}) and (\ref{g_appr_appr}). On one hand, the enhancement effect on vacuum effect depends only on the expression of the refractive index $n_0$ which, in the present framework, is not modified by the presence of collisional effects. On the other hand, the collisional effects induce a reduction of the wave amplitude and the arising of a frequency chirping. Actually, the attenuation of the wave amplitude does not change the conclusions in Sec. \ref{Setup}. It is only understood that the values of the amplitudes $E'_{0,1}$ and $E'_{0,2}$ introduced in Eq. (\ref{E}) will be different from those used above. Instead, the total polarization rotation angle $\Delta\Phi(z)$ contains not only the quantity $\Delta\phi(z)$ already considered  before [see Eq. (\ref{Delta_phi})] but it also receives a chirping contribution $\Delta\phi_{\text{ch}}(z)$ that is given by
\begin{equation}
\label{Delta_phi_ch} 
\Delta\phi_{\text{ch}}(z)=\frac{\omega^2z^2}{2}\left[g''_2(0)n_{0,2}^2-g''_1(0)n_{0,1}^2\right]
\end{equation}
and it is proportional to $z^2$. In this way, in order to isolate completely the VPEs we have to exclude the chirping term from the polarization rotation angle. This can be achieved, at least, in two ways. For example, we could consider so strong fields that $|\Delta\phi_{\text{ch}}(z)|\ll \delta_{\text{vac}}\omega z/(n_{0,1}+n_{0,2})$. We have shown numerically that this strong inequality is fulfilled at laser intensities $I_0\sim 10^{26}\;\text{W/cm$^2$}$ for distances $z$ of the order of ten laser wavelength in plasma. Another way is to perform two phase measurements $\Delta\Phi(z_A)$ and $\Delta\Phi(z_B)$ differing only in the plasma lengths $z_A$ and $z_B$. In this case, from Eqs. (\ref{Delta_phi}) and (\ref{Delta_phi_ch}) we have $\Delta\phi(z_A)z_B=\Delta\phi(z_B)z_A$ and $\Delta\phi_{\text{ch}}(z_A)z_B^2=\Delta\phi_{\text{ch}}(z_B)z_A^2$. Finally, these relations allow us to express the linear phase difference $\Delta\phi(z_A)$ in terms of the experimental quantities $\Delta\Phi(z_A)=\Delta\phi(z_A)+\Delta\phi_{\text{ch}}(z_A)$ and $\Delta\Phi(z_B)=\Delta\phi(z_B)+\Delta\phi_{\text{ch}}(z_B)$ as
\begin{equation}
\Delta\phi(z_A)=\frac{\Delta\Phi(z_A)z_B^2-\Delta\Phi(z_B)z_A^2}{z_B(z_B-z_A)}.
\end{equation}
%
%
%
\section{Summary and conclusions}
\label{Conclusions}
In this paper we have calculated for the first time the dispersive effects of vacuum polarization on the propagation of a strong monochromatic wave through a cold collisional plasma. We have considered the easiest situation of the propagation of a circularly polarized wave in order to describe the new features introduced by the vacuum polarization and to obtain analytical estimates of the vacuum polarization effects (VPEs). The most important result is that, in the presence of a plasma, the VPEs are strongly enhanced with respect to those in pure vacuum. As we mentioned, the physical reason of this enhancement is connected with the singular dielectric behaviour of a plasma near the plasma frequency: the pure plasma refractive index becomes much smaller than unity for a wave with frequency only slightly larger than the plasma frequency. We have also proposed an experimental setup to measure these VPEs by making two strong fields with the same frequency and different intensities to interfere after crossing two plasmas with different electron and ion densities.

It is also worth noting here that, independently on VPEs, the implicit analytical solution we have found in Eqs. (\ref{h_p}) and (\ref{g_appr}) holds under the very general conditions (\ref{cond_1}) and (\ref{cond_2}) and it can be a useful analytical tool to describe the propagation of a strong wave through a relativistic collisional plasma. 

We have already pointed out in Sect. \ref{Setup} the practical and conceptual difficulties of our approach. Nevertheless, our aim is to stress that at the high laser intensities feasible in the near future the inclusion of the VPEs in studying the wave propagation through a plasma is, on one hand, required for the sake of consistency and, moreover, it can open other and new possibilities to detect the VPEs themselves.
%
%
%
\begin{figure}
\begin{center}
\includegraphics[width=14cm]{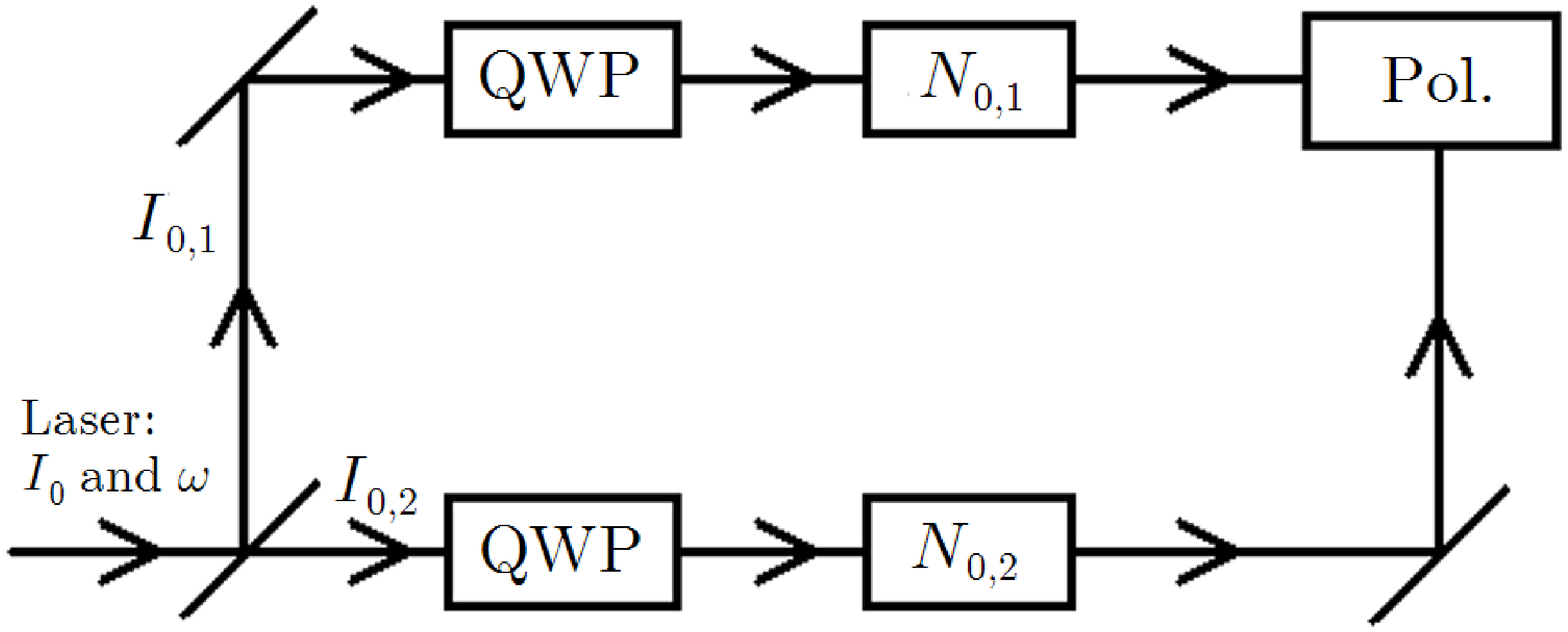}
\end{center}
\caption{Schematic experimental setup described in Sec. \ref{Setup}. The ``QWP'' boxes represent two quarter-wave plates, the ``$N_{0,1}$'' and ``$N_{0,2}$'' boxes represent the two plasmas with electron densities $N_{0,1}$ and $N_{0,2}$ and the ``Pol.'' box the polarimeter.}
\end{figure}
\begin{figure}
\begin{center}
\includegraphics[width=14cm]{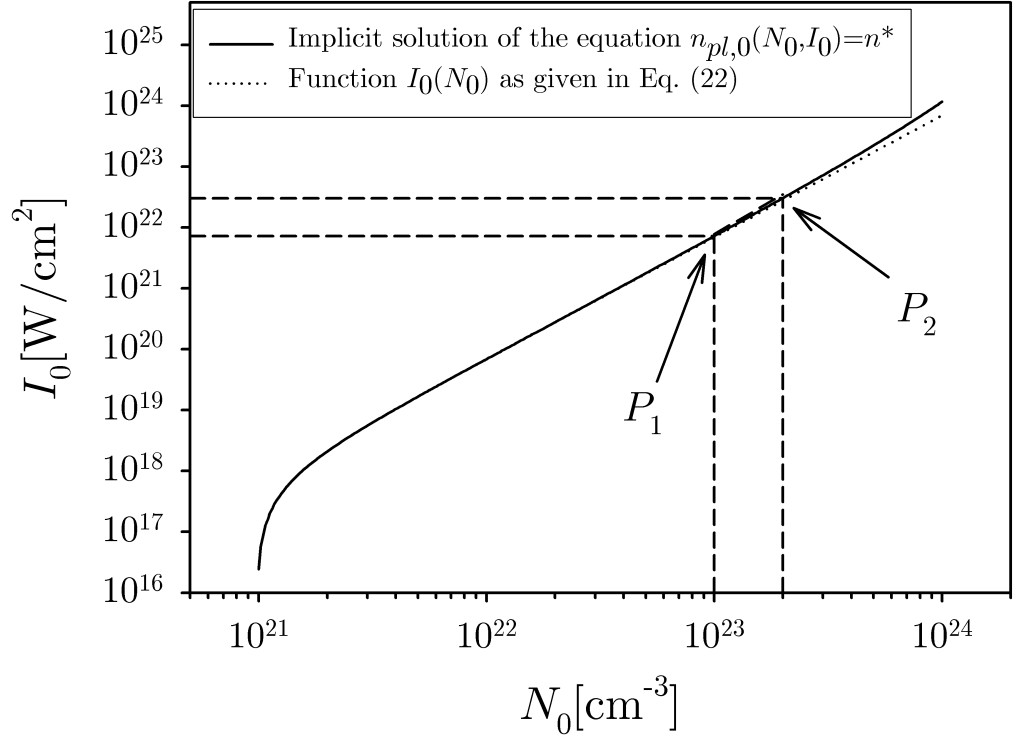}
\end{center}
\caption{Log-Log plot of the exact numerical solution of the equation $n_{pl,0}(N_0,I_0)=n^*$ (continuous line) and the approximated analytical solution given in Eq. (\ref{I_N}) (dotted line). The value of $n^*$ has been set $n^*=5\times 10^{-2}$. The two points $P_1=(N_{0,1},I_{0,1})$ and $P_2=(N_{0,2},I_{0,2})$, correspond to the situation in the numerical example given below Eq. (\ref{I_N}).}
\end{figure}
\begin{figure}
\begin{center}
\includegraphics[width=14cm]{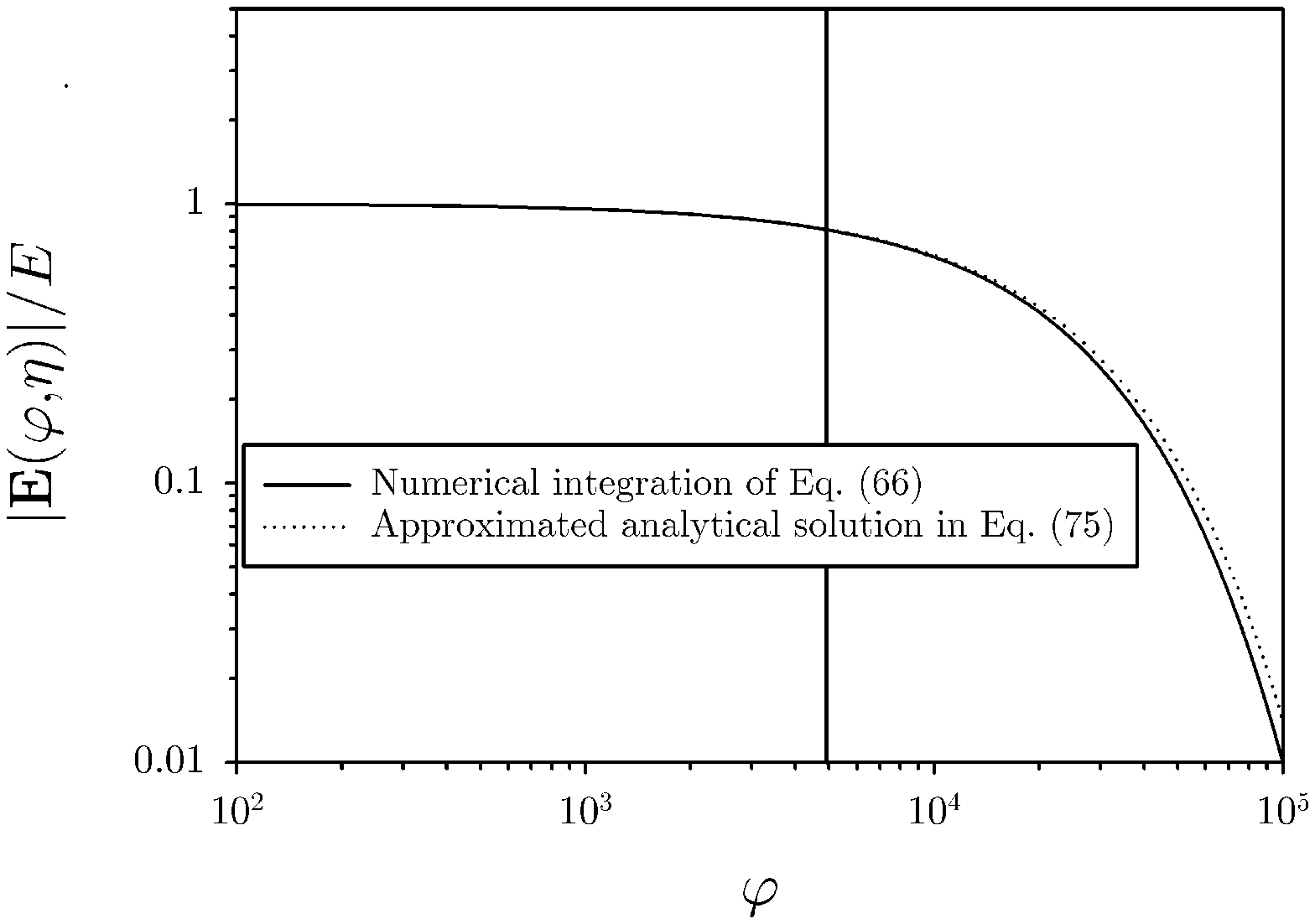}
\end{center}
\caption{Log-log plot of the typical behaviour of the amplitude of the electric field (\ref{E_fin_coll}) in unit of $E$ as a function of the adimensional parameter $\varphi$ obtained by integrating Eq. (\ref{h_p}) (continuous line) and by using the approximated solution in Eq. (\ref{h_appr_appr}) (dotted line). From Eq. (\ref{E_fin_coll}) one sees that $|\mathbf{E}(\varphi,\eta)|$ depends actually only on $\varphi$. The laser frequency $\omega$ has been set equal to $1.1\sqrt{\omega^2_{pl,e}+\omega^2_{pl,i}}$}. The vertical line corresponds to the value $\varphi_0\approx 1/(10\delta_{\nu})$ up to which the approximated solution is theoretically predicted to be accurate.
\end{figure}


\begin{thebibliography}{99}
\bibitem{Footnote_2} Since we are considering also the VPEs perturbatively up to first order in $\delta_{\text{vac}}$ it would be more correct to set also $\delta_{\text{vac}}=0$. Nevertheless, the VPEs do not affect the zero-order Lorentz factors $\gamma_{0,\lambda}$ that are the only dynamical quantities entering the collision frequency because $p_0=m_e\sqrt{\gamma^2_{0,_e}-1}$.
\end{thebibliography}
\end{document}